\documentstyle[preprint,aps]{revtex}
\begin{document}
\title{Quantum equivalent of the Bertrand's theorem}
\author{N. Gurappa, Prasanta K. Panigrahi and T. Soloman Raju}
%\thanks{panisp@uohyd.ernet.in} 
\address{School of Physics, 
University of Hyderabad,
Hyderabad,\\ Andhra Pradesh,
500 046 INDIA.}
\maketitle

\begin{abstract} 
A procedure for constructing bound state potentials is given. We show that,
under the natural conditions imposed on a radial eigenvalue problem, the {\it
only} special cases of the general central potential, which are exactly
solvable and have infinite number of energy eigenvalues, are the Coulomb and
harmonic oscillator potentials.
\end{abstract}
\draft

\newpage

The celebrated Bertrand's theorem in classical mechanics states that {\it the
only central forces that result in closed orbits for all bound particles are
the inverse square law and Hooke's law} \cite{ber,gold}. The extension of the
Bertrand's theorem to spherical geometry led again to two potentials for
which closed orbits exist \cite{higgs}. In the limit of vanishing curvature,
these potentials reduced to the oscillator and Coulomb potentials. As is
well known, these two potentials are also special in quantum mechanics.
These are the only ones, which can be solved exactly in arbitrary dimensions
and have infinite number of bound states. From the symmetry point of view,
the degeneracy structure of these potentials admit a Lie group larger than
the $SO(d)$ group for a non-relativistic particle moving in $d$-dimensional
Euclidean space \cite{bac,ste}. For the Coulomb problem, the group is
$SO(d+1)$ for bound states and $SO(d, 1)$ for scattering states. For harmonic
oscillator, the corresponding group is $SU(d)$. In case of spherical
geometry, the eigenvalue problem can be solved exactly only for the two, 
above mentioned potentials \cite{higgs,lak}. These are the only central
potentials for which the corresponding Schr\"odinger equations can be
factorized to yield both the energy and angular momentum raising and lowering
operators \cite{sch,pr,fac1,fac2}. These two potentials are also special from
the semi-classical point of view \cite{iva}. Recently, it has been shown that,
starting from a suitably gauged action, these two potentials can be derived
in lower dimensions by appropriate gauge fixing \cite{bars}. However, unlike
the classical case where a general central potential, under the twin
constraints of bound and closed orbits, led to the Coulomb and oscillator
potentials; in quantum mechanics, the natural conditions imposed on a radial
eigenvalue problem, in a given dimension, have not yielded these two
potentials as unique choices.

In this paper, we construct a family of potentials, dependent on a parameter
$\alpha$, keeping in mind the two important features of the quantum
mechanical bound state problems, {\it i.e.}, the discreteness of their
eigenspectra and the polynomial nature of the corresponding normalizable
wavefunctions. Interestingly, under these general conditions, there are only
two central potentials which are exactly solvable and have infinite number of
bound states. These two unique potentials, appearing for the two values of
$\alpha$, one and two, are the Coulomb and the harmonic oscillator
potentials, respectively. This quantum mechanical situation is an exact
analogue of the Bertrand's theorem in classical mechanics.

We begin with a Fock space spanned by the monomials, $\rho^n$; where, $n =
0,1, 2, \cdots$. The most general diagonal operator which has a well defined
action on this Fock space can be written as $\sum_{k = - \infty}^\infty C_k
D^k$; here, $D \equiv \rho \frac{d}{d \rho}$ and $C_k$'s are constants. The
two special cases  corresponding to, (i) all the $C_k$'s being zero except
$C_1 = 1$ and $C_0 = \epsilon$ (modulo an overall scaling factor) and (ii)
all the $C_k$'s being zero, except $C_2 = 1$, $C_1 = \beta$, and $C_0 =
\delta$, can be unambiguously mapped to the Schr\"odinger equation by a
series of similarity transformations. The first case leads to a class of
potentials, dependent on a parameter $\alpha$. It is explicitly shown that,
the only two exactly solvable radial potentials, having infinite number of
energy eigenvalues, are the harmonic and Coulomb potentials.  Our method
naturally gives their respective eigenfunctions. Further possibilities of
solvable potentials are analyzed by performing a point canonical
transformation. It is found that, no other exactly solvable central potential
results and Morse potential arises as a conditionally exactly solvable case
of this general potential. We then construct the second class of potentials
and point out that, unlike the first, this case can not be solved as a radial
eigenvalue problem, for infinite number of levels, for all the allowed values
of $\alpha$ and for the various possible choices of the other parameters
appearing in it.

The choice (i), mentioned above, yields,
\begin{equation} \label{eu}
\left(\rho \frac{d}{d \rho} + \epsilon \right) \eta(\rho) = 0 \qquad;
\end{equation}
where, $\eta(\rho) = \rho^{- \epsilon}$. 

Performing a similarity transformation (ST): $\phi(\rho) =
\exp\{\hat{A}/\alpha\} \eta(\rho)$, where, $\hat{A} \equiv a \rho^{2 -
\alpha} \frac{d^2}{d \rho^2} + b  \rho^{1 - \alpha} \frac{d}{d \rho} + c
\rho^{- \alpha}$, $\alpha$, $a, b$ and $c$ being arbitrary parameters, one
gets 
\begin{equation} \label{bc}
(\rho \frac{d}{d \rho} + a \rho^{2 - \alpha} \frac{d^2}{d \rho^2} 
+ b \rho^{1 - \alpha} \frac{d}{d \rho} + c \rho^{- \alpha} + \epsilon) \phi =
0 \qquad. 
\end{equation}
The above choice of the ST is motivated from our desire to map (\ref{eu}) to
the Schr\"odinger equation and also to have a closed form expression for the
second order differential equation.

Choosing $\phi = \rho^{\alpha - 2} \chi$, it is straightforward to check
that,
\begin{eqnarray} \label{chi}
\frac{d^2 \chi}{d \rho^2}&& + \frac{1}{a} \left(\rho^{\alpha - 1} + (b - 4 a
+ 2 a \alpha) \rho^{-1}\right) \frac{d \chi}{d \rho} \nonumber\\
&&+ \frac{1}{a}
\left((\epsilon + \alpha - 2) \rho^{\alpha - 2} + (a \alpha^2 + (b - 5 a)
\alpha + 6 a - 2 b + c) \rho^{- 2}\right) \chi = 0 \qquad.
\end{eqnarray}
For the purpose of clarity, we will cast (\ref{chi}) as a radial eigenvalue
equation in three dimensions; it can be easily seen that our method
generalizes to $d$-dimensions. Taking $\psi(\rho) = S(\rho)
\chi(\rho)$, one gets
\begin{equation} \label{sch}
\left(\frac{d^2}{d r^2} + \frac{2}{r} \frac{d}{d r} - \frac{l(l + 
1)}{r^2} + \frac{2 m}{{\hbar}^2} (E - V(r)) \right)
\psi = 0 \qquad,
\end{equation}
where, the radial coordinate $r = \lambda \rho$, $\lambda$ being an
appropriate length scale. Here, $S(\rho) = \rho^{A_0} \exp\{\rho^\alpha/(2 a
\alpha)\}$, $A_0 \equiv (b - 6 a + 2 a \alpha)/(2 a)$ and $l$ is the
conventional angular momentum quantum number.

The explicit form of the potential is 
\begin{equation} \label{gcp}
V(r) = E + g_1 r^{2 (\alpha - 1)} + g_2 r^{\alpha - 2} + g_3 r^{- 2} \qquad,
\end{equation}
where, $g_1 = \tilde{g}_1 \lambda^{2 (1 - \alpha)}$, $g_2 = \tilde{g}_2
\lambda^{2 - \alpha}$, $g_3 = \tilde{g}_3 \lambda^2$, $\tilde{g}_1 =
{\hbar}^2/(8 m a^2 \lambda^2)$, $\tilde{g}_2 = {\hbar}^2 (2 A_0 - \alpha + 5
-2 \epsilon)/(4 m a \lambda^2)$ and $\tilde{g}_3 = {\hbar}^2 \{A_0 (A_0 + 1) -
l (l + 1) - [a (2 - \alpha) (3 - \alpha) + b \alpha - 2 b + c]/a\}/(2 m
\lambda^2)$.  

The corresponding unnormalized eigenfunctions, obtained through inverse
similarity transformations, are
\begin{equation}
\psi(\rho) = \rho^{(\pm \sqrt{\Delta} - 1)/2} \exp\left\{\rho^{\alpha}/(2 a
\alpha) \right\} L_n^{\pm \sqrt{\Delta}/\alpha}\left(- \rho^{\alpha}/(a \alpha)
\right) \qquad, 
\end{equation}
where, $\Delta \equiv (1 - b/a)^2 - 4 c/a$. Sign of $\sqrt{\Delta}$ should be
chosen such that the resulting wavefunctions are normalizable. The Laguerre
polynomial, $L_n^{\pm \sqrt{\Delta}/\alpha}$, is obtained by demanding that
the $(n + 1)th$ term in $\exp\{\hat{A}/\alpha\} \rho^{- \epsilon_n}$ is
equal to zero\cite{ak}; {\it i.e.}, $\hat{A}^{n + 1} \rho^{- \epsilon_n} =
0$. This gives  
\begin{equation} \label{eev}
\epsilon_n^{\pm} = - \alpha n - (1/2) (1 - b/a) \pm
(1/2)\sqrt{\Delta}\qquad. 
\end{equation}

Before proceeding to study the exact solvability of the special cases of the
bound state problem obtained in (\ref{gcp}), it is worth repeating the
corresponding scenario in classical mechanics. In classical case, one demands
the bounded motion of a test particle, subjected to a general central force
of the form $F(r) = - \kappa r^{\beta^2 - 3}$, to trace a {\it closed orbit};
this leads to two possible values for $\beta$, one and two\cite{gold}. These
values give rise to the well known inverse-square and Hooke's laws,
respectively. Interestingly, our general construction leads to a central
potential, governed by a real parameter $\alpha$. Much akin to the classical
case, {\it only} for $\alpha = 1$ and $2$, as will be shown below, this
potential can be solved for the complete set of energy eigenvalues. This is
a quantum analogue of the classical Bertrand's theorem.

The general central
potential in (\ref{gcp}) has free parameters $g_1$ or $g_2$ for $\alpha 
= 1$ or $2$: one can then impose $E + g_i = 0$, for $i = 1, 2$, in order
that the potential is independent of constant terms and the
corresponding Schr\"odinger equation is exactly solvable for all the energy
eigenvalues.

{\it Case I}: $\alpha = 1$.

To cast the eigenvalue problem in the standard radial form, {\it i.e.}, to
make $g_3 = 0$, we choose $a = 1/(2 \sigma)$, $b = 2/\sigma$ and $c = [2 -
l(l + 1)]/(2 \sigma)$ with $\sigma^2 = - 2 m \lambda^2 E/{\hbar}^2$.
We now demand that $E +g_1 = 0$. (\ref{gcp}) then reduces to the Coulomb
potential, 
$$ V(r) = \frac{g_2}{r} \qquad.$$

Choosing $g_2 = - e^2/(4 \pi \epsilon_0)$, $\epsilon_n^-$ in (\ref{eev})
gives the energy eigenvalues as 
$$ E = -  g_1 = - \frac{m e^4}{32 \pi^2 \epsilon_0^2 {\hbar}^2} \frac{1}{(n +
l + 1)^2} \qquad.$$

{\it Case II}: $\alpha = 2$.

Akin to the previous case, choosing $a = 1/(2 \sigma)$, $b = 1/\sigma$ and $c
= - l (l + 1)/(2 \sigma)$ with $\sigma = - m \omega \lambda^2/\hbar$, one can
check that $g_3 = 0$. Further demanding $E + g_2 = 0$, the potential reduces
to the harmonic oscillator potential: 
$$V(r) = g_1 r^2 \qquad.$$

For, $g_1 = m \omega^2/2$, the eigenspectra is 
$$E = -  g_2 = \hbar \omega (2 n + l + 3/2) \qquad.$$
It is worth pointing out that only for these two values of $\alpha$, one can
make the potential independent of constants and hence obtain the full
spectrum. We would like to add that, the above two cases can also be solved
for $g_3 \ne 0$\cite{lan,mpl,fac3}. For values of $\alpha$ different from
one and two, the potential can be solved for $E = 0$, but for different
values of the angular momentum quantum number l. This situation is
similar to the classical scenario, where, some values of $\beta$ other than
$1$ and $2$ may give rise to bounded motion, which is not closed.

At this moment, one is naturally curious to see whether there exists the
possibility of exactly solvable models emerging, under a point canonical
transformation (PCT) $\rho \rightarrow f(\rho)$ when $\alpha \ne 1,
2$\cite{de,pkp}. Performing the PCT, (\ref{bc}) can be brought to the form
(\ref{sch}), with a potential given by
\begin{eqnarray} \label{pct}
\frac{2 m \lambda^2}{{\hbar}^2} V(\lambda \rho) &=& \frac{2 m
\lambda^2}{{\hbar}^2} E + \frac{1}{4 a^2} (f^\prime f^{\alpha - 1})^2 
+ \frac{1}{2 a^2} [b + a (\alpha - 1 - 2 \epsilon)]
(f^\prime)^2 f^{\alpha - 2} \nonumber \\
&&+ \left[\frac{b}{2 a} \left(\frac{b}{2 a} -
1\right) - \frac{c}{a}\right] \left(\frac{f^\prime}{f}\right)^2 - \frac{l (l +
1)}{\rho^2} \nonumber\\
&&+ \frac{3}{4} \left(\frac{f^{\prime \prime}}{f}\right)^2 
- \frac{1}{2} \frac{f^{\prime \prime \prime}}{f^\prime}
- 2 \frac{f^{\prime \prime}}{f f^\prime} 
+ (2 - \alpha) \frac{f^\prime f^{\prime
\prime}}{f^2} + 2 \left(\frac{f^{\prime \prime}}{f^\prime}\right)^2 \qquad,
\end{eqnarray}
where, $f^\prime \equiv df/d\rho$.

Different choices of $f$ will give rise to different potentials; however, all
of them may not be exactly solvable. For example the potential in (\ref{pct}) can
be made independent of constants, even for continuous values of
$\alpha$, if one chooses $f(\rho) = e^\rho$:
\begin{eqnarray}
\frac{2 m \lambda^2}{{\hbar}^2} V(\lambda \rho) = \frac{1}{4 a^2} e^{ 2
\alpha \rho} + \frac{1}{2 a^2} [b + a (\alpha - 1 - 2 \epsilon)] e^{\alpha
\rho} - 2 e^{- \rho} - \frac{l (l + 1)}{\rho^2} \qquad.
\end{eqnarray}
Since, the above potential contains the term `$- l (l + 1) \rho^{- 2}$', one
can only solve the eigenvalue equation for a fixed `$l$'. This is the case of
the conditionally exactly solvable potential encountered in Ref.\
\onlinecite{mpl}. Physically this implies that, the potential is exactly
solvable only in one dimension, {\it i.e}, without the centrifugal term.
The corresponding energy eigenvalues are independent of
$l$; they can be obtained from $$\frac{2 m \lambda^2}{{\hbar}^2} E +
\frac{b}{2 a} \left(\frac{b}{2 a} - 1\right) - \frac{c}{a} - \alpha + 17/4 =
0 \qquad,$$ and the unnormalized eigenfunctions are 
$$\psi = \frac{1}{\rho} f^{(\pm \sqrt{\Delta} + 1)/2}
\exp\left\{f^{\alpha}/(2 a \alpha)\right\} L_n^{\pm
\sqrt{\Delta}/\alpha}\left(- f^{\alpha}/(a \alpha)\right) \qquad.$$ 
One can easily see that, the Morse potential emerges as a special case for
$\alpha = - 1$.
%*****************************************************************************

Now, we proceed to find the second class of potentials starting from
\begin{equation}
\left(D^2 + \beta D + \delta \right) \bar{\eta}(\rho)  = 0 \qquad.
\end{equation}
Writing $\bar{\chi}(\rho) = \exp\{- \gamma \hat{O}\} \bar{\eta}(\rho)$,
where $\hat{O} \equiv a \rho^\alpha \frac{d}{d \rho} + b \rho^{\alpha - 1}$,
one gets 
\begin{equation} \label{F}
\left(F_1(\rho) \frac{d^2}{d \rho^2} + F_2(\rho) \frac{d}{d \rho} + F_3(\rho) 
+ \delta\right)\bar{\chi} = 0 \quad, 
\end{equation}
where, $F_1(\rho) = A_1 \rho^2 (\rho^{\alpha - 1} + A_2)^2$, $F_2(\rho) = B_1
\rho (\rho^{\alpha - 1} + B_2)^2 + B_3$ and $F_3(\rho) = C_1 (\rho^{\alpha
- 1} + C_2)^2 + C_3$. The coefficients are given by $A_1 = A_2^{- 2}$,
$A_2^{-1} = (\alpha - 1) \gamma a$, $B_1 = (\alpha a + 2 b)/(a A_2^2)$, $B_2
= [2 b + a (3 + \beta - \gamma)]/(2 a A_2 B_1)$, $B_3 = 1 + \beta - B_1
B_2^2$, $C_1 = b (a \alpha - a + b)/(a^2 A_2^2)$, $C_2 = b (\beta - \gamma +
1)/(2 a A_2 C_1)$ and $C_3 = - C_1 C_2^2/b$. Here, if one chooses $\hat{A}$
instead of $\hat{O}$, then the resulting equation  will contain
still more higher derivative terms and can not be straightforwardly mapped to
a Schr\"odinger equation.

Now, writing $\bar{\chi}(\rho) = \left(F_1(\rho) S(\rho)\right)^{-1}
\bar{\psi}(\rho)$; where, $S(\rho)$ can be obtained from
$$\frac{1}{S} \frac{d S}{d \rho} = \frac{F_2(\rho) - 2 F_1^{\prime}(\rho)}{2
F_1(\rho)} - \frac{1}{\rho} \qquad,$$
it can be checked that, $\bar{\psi}$ obeys the Schr\"odinger equation with a
potential, 
\begin{eqnarray}
\tilde{V}(\lambda \rho) &=& \tilde{E} + \frac{[B_1 \rho (\rho^{\alpha - 1} +
B_2)^2 + B_3][D_1 \rho (\rho^{\alpha - 1} + D_2)^2 + D_3]}{4 A_1^2 \rho^4
(\rho^{\alpha - 1} + A_2)^4} \nonumber\\
&&+ \frac{D_1 (\rho^{\alpha - 1} + D_2) [(2 \alpha - 1) \rho^{\alpha - 1} +
D_2]}{A_1 \rho^2 (\rho^{\alpha - 1} + A_2)^2} \nonumber\\
&&+ \frac{8}{\rho^2} \left(\frac{\alpha \rho^{\alpha - 1} + A_2}{\rho^{\alpha -
1} + A_2}\right)^2 \left(\frac{1}{\rho} + \frac{\alpha - 1}{\rho^{\alpha - 1}
+ A_2} + \frac{\alpha (\alpha - 1)}{\alpha \rho^{\alpha - 1} + A_2} \right)
\nonumber\\ 
&&- \frac{[C_1 (\rho^{\alpha - 1} + A_2)^2 + C_3 + \delta]}{A_1 \rho^2
(\rho^{\alpha - 1} + A_2)^2} - \frac{l (l + 1)}{\rho^2} \qquad.
\end{eqnarray}
Here, $\tilde{V}(\rho ) \equiv \frac{2 m \lambda^2}{{\hbar}^2} V(\lambda
\rho)$, $\tilde E \equiv \frac{2 m \lambda^2}{{\hbar}^2} E$, $D_1 = (2 b - 3
\alpha a)/(a A_2^2)$, $D_2 = [2 b + a (\beta - \gamma - 4 \alpha - 1)]/(2 A_2
D_1)$ and $D_3 = 1 + \beta - 4/D_1 - D_1 D_2^2$. 

Unlike the first class, these potentials can not be made independent of
constants as a general radial problem,  for any special values of $\alpha.$ Hence, the possibility of
obtaining infinite number of levels, in arbitrary dimensions, does not arise here. However, it can be
solved for the $E=0$ state, for different values of $l$.

At this point, we would like to add that, there have been many works in
the literature, where point canonical transformation is used after writing 
$\psi(x)=f(x)F[g(x)] $, for mapping 
the Schr\"odinger equation to the confluent hypergeometric or
hypergeometric equations \cite{ecg,nat,gin,khare}. Hence, from the known 
solutions of these equations, one is able to obtain solvable potentials in the
Schr\"odinger eigenvalue problem, after imposing certain restrictions on
f and g.
In our case, we have started with the diagonal operators 
$D \equiv \rho \frac{d}{d \rho}$ 
and $D^{2}+\alpha{D}$  acting on the space of monomials $\rho^n.$
By inverse transformation, we, not only obtained the Schr\"odinger
eigenvalue equation, but also the connection of the respective
eigenfunctions with the monomials. In this approach, the special nature of
the Coulomb and oscillator potentials, {\it i.e.}, exact solvability with
infinite number of levels in arbitrary dimensions, came out naturally
without taking recourse to PCT. 

We would also like to point out that, a number of
Calogero-Sutherland  type  interacting, many-body Hamiltonians, both in one
and higher dimensions, can be mapped to the Euler operator 
$\sum x_i\frac{\partial}{\partial{x_i}}+c$,  through similarity 
transformations\cite{prb}. In fact, our partial motivation behind this work 
is due to a theorem established in Ref.\ \onlinecite{prb}:

All $D$ dimensional $N$ particle Hamiltonians, which can be brought through a
suitable transformation to the generalized form: $\tilde H = \sum_{l=1}^D
\sum_{i=1}^N x_i^{(l)} \frac{\partial}{\partial x_i^{(l)}} + \epsilon + \hat
A$ can also be mapped to $\sum_{l=1}^D \sum_{i=1}^N x_i^{(l)}
\frac{\partial}{\partial x_i^{(l)}} + \epsilon$ by $\exp\{- d^{-1} \hat A\}$;
where, the operator $\hat A$ is any homogeneous function of
$\frac{\partial}{\partial x_i^{(l)}}$ and $x_i^{(l)}$ with degree $d$ and
$\epsilon$ is a constant. For the normalizability of the wave functions, one
needs to check that the action of $\exp\{- d^{-1} \hat A\}$ on an appropriate
linear combination of the eigenstates of $\sum_{l=1}^D \sum_{i=1}^N x_i^{(l)}
\frac{\partial}{\partial x_i^{(l)}}$ yields a polynomial solution. 
It will be of great
interest to apply the method employed here to the many-body systems and 
examine carefully, how far the results obtained for the single particle case
generalizes to the N-particle systems. It is worth mentioning here that, apart
from the Cologero-Sutherland model, which has recently been mapped to harmonic
oscillators\cite{prb}, the other many-particle exactly solvable system is of 
Coulombic type\cite{ak}.

%\newpage
The authors would like to acknowledge useful discussions with Profs. V.
Srinivasan and S. Chaturvedi. N.G thanks U.G.C (India) for the financial support through the
S.R.F scheme.


\begin{thebibliography}{99}

\bibitem{ber} J. Bertrand, Comptes Rendus {\bf 77}, 849 (1873).

\bibitem{gold} H. Goldstein, Classical Mechanics, 2nd ed., (Addison-Wesley,
New York, 1980).

\bibitem{higgs} P.W. Higgs, J. Phys. {\bf A}: Math. Gen., {\bf 12}, 309
(1979). 

\bibitem{bac} H. Bacry, H. Ruegg and J.M. Souriau, Commun. Math. Phys. {\bf
3}, 323 (1966).

\bibitem{ste} P. Stehle and M.Y. Han, Phys. Rev. {\bf 159}, 1076 (1967).

\bibitem{lak} M. Lakshmanan and K. Eswaran, J. Phys. {\bf A}: Math. Gen. {\bf
8}, 1658 (1975).

\bibitem{sch} E. Schr\"odinger, Proc. R. Ir. Acad. {\bf 46 A}, 9 (1940).

\bibitem{pr} F. Cooper, A. Khare and U. Sukhatme, Phys. Rep. {\bf 251}, 268
(1995) and references therein. 

\bibitem{fac1} J.D. Newmarch and R.M. Golding, Am. J. Phys. {\bf 46}, 658
(1978). 

\bibitem{fac2} Y.F. Liu, Y.A. Lei and J.Y. Zeng, Phys. Lett. {\bf A 231}, 9
(1997). 

\bibitem{iva} I.A. Ivanov, J. Phys. {\bf A}: Math. Gen., {\bf 29}, 3203
(1996). 

\bibitem{bars} I. Bars, pre-print no. USC-98/HEP-B2, hep-th/9804028.

\bibitem{ak} N. Gurappa, A. Khare and P.K. Panigrahi, Phys. Lett. {\bf A
224}, 467 (1998).

\bibitem{lan} L. Landau and E.M. Lifshitz, Quantum Mechanica:
Non-Relativistic Theory (Oxford Univ. Press, 1975).

\bibitem{mpl} N. Gurappa, C.N. Kumar and P.K. Panigrahi, Mod. Phys. Lett.
{\bf A 11}, 1737 (1996). 

\bibitem{fac3} Z.B. Wu and J.Y. Zeng, J. Math. Phys. {\bf 39}, 5253 (1998).

\bibitem{de} R. De, R. Dutt and U. Sukhatme, J. Phys. {\bf A}: Math. Gen.
{\bf 25}, L843 (1992).

\bibitem{pkp} A. Gangopadhyaya, P. Panigrahi and U. Sukhatme, Helv. Phys.
Acta. {\bf 67}, 363 (1994).

\bibitem{ecg} A. Bhattacharjie and E.C.G. Sudarshan, Nuovo Cimento {\bf 25},
864 (1962).

\bibitem{nat} G.A. Natanzon, Teor. Mat. Fiz. {\bf 38}, 146 (1979).

\bibitem{gin} J.N. Ginocchio, Ann. Phys. (N.Y.) {\bf 152}, 203 (1984).

\bibitem{khare} F. Cooper, A. Khare and U. Sukhatme, Phys. Rep. {\bf 251},
267 (1995), and references therein.

\bibitem{prb} N. Gurappa and P.K. Panigrahi, Phys. Rev. {\bf B}, R2490
(1999), and references therein.

\bibitem{ak} A. Khare, J. Phys. {\bf A}: Math. Gen. {\bf 29}, L45 (1996).

\end{thebibliography}
\end{document}